\documentclass{ws-procs975x65}

\begin{document}

\title{\uppercase{Experimentally testing asymptotically safe quantum gravity with photon-photon scattering}}

\author{\uppercase{Astrid Eichhorn}}

\address{Perimeter Institute for Theoretical Physics, 31 Caroline Street N, Waterloo, N2L 2Y5, Ontario, Canada
$^*$E-mail: aeichhorn@perimeterinstitute.ca}

\begin{abstract}
Matter-quantum gravity interactions can be used for direct and also indirect experimental tests of quantum gravity. We focus on photon-photon scattering in asymptotically safe gravity as a direct test of the small-scale structure of spacetime, and discuss how near-future experiments can probe asymptotic safety in a setting with large extra dimensions.
\end{abstract}

\bodymatter
\section{Introduction}
In quantum gravity research, we strive to answer the question "What is spacetime like on very small scales?". More precisely, this can be rephrased as "What is the microscopic dynamics of gravity?", "What are the fundamental symmetries?", "Is spacetime continuous or somehow discrete?", "What is the dimensionality of spacetime on small scales?". To answer these, each approach to quantum gravity is founded upon assumptions. 
Ultimately, these should be tested in experiments. Here, matter-quantum gravity interactions provide us with two possibilities:

1) consistency tests (indirect): Here we can check, whether a particular UV completion for gravity is compatible with the observed properties of the standard model of particle physics, such as the number of fermions, scalars and gauge bosons, the global symmetries, the observed mass scales, etc.

2) direct tests: Effects such as graviton-exchange in scattering processes, can be used to study the small-scale structure of spacetime. In particular, the dimensionality of spacetime on small scales can be tested at  near-future experiments.

For examples of the first option, where the consistency of asymptotically safe quantum gravity, as well as other models for quantum gravity, with the existence of light fermions in our universe is tested, see \cite{Eichhorn:2011ec}. Within the truncation of the full Renormalization Group (RG) flow employed in this work, asymptotic safety is compatible with fermion masses much below the Planck mass, whereas restrictions are placed on the parameter space for other quantum gravity models.

\section{Testing the small-scale structure of spacetime with high-intensity lasers}
High-intensity lasers are subject to quickly advancing experimental research, and have the potential to provide future particle accelerators that could replace conventional accelerators and reach higher energies over much shorter acceleration paths, albeit reaching high luminosities remains a challenge. Once high-energy electrons are available, either from high-intensity lasers or conventional linear accelerators, the energy can be transferred to photons by Compton-backscattering, thus enabling high-energy photon-photon scattering. This provides for a possible window into the quantum-gravity regime \cite{Dobrich:2012nv}: Since there is no microscopic photon-photon interaction in the standard model, the cross-section is dominated solely by loop effects.  Due to the tree-level graviton-exchange in the $s$, $t$ and $u$ channels, the quantum gravity effect is easier to access here than in processes which have a large standard model background, such as, e.g., processes at the LHC, see, e.g. \cite{Litim:2007iu}.

Here, we will focus on asymptotic safety as a UV completion for gravity, i.e. a quantum field theory of the metric, which is extendible up to arbitrarily high energies due to the existence of an interacting fixed point of the RG.
See, e.g. \cite{Reuter:2012id} for reviews.
In order to detect experimental signatures of asymptotic safety, we have to include the effect of a momentum-scale dependent Newton coupling, which encodes the leading-order non-perturbative physics underlying the asymptotic-safety scenario. 
The dimensionful Newton coupling $G_N = {\rm const}$ on a large range of scales. This behavior characterizes 
the classical regime of the theory, required for the consistency with experiment, and has been shown to exist in truncated RG flow studies \cite{Reuter:2004nx}. 
The second property is special in the asymptotic-safety scenario:  A UV completion with the help of an interacting fixed point requires that $G(k) \rightarrow G_{\ast}$ for 
$k \rightarrow \infty$, where $G(k)=G_N(k)k^{d-2}$ is the dimensionless Newton coupling and $k$ the RG scale. Thus we model the scale-dependence in $d$ spacetime dimensions in the following way:
\begin{equation}
G_N(k) = \theta(k_{\rm tr}^2 - k^2) G_{\rm Newton}+ \theta(k^2 - k_{\rm tr}^2) G_{\rm Newton} \frac{k_{\rm tr}^{d-2}}{k^{d-2}}\label{runningG},
\end{equation}
where $G_{\rm Newton}$ is the constant dimensionful value for the Newton coupling measured in the infrared. 
Here the transition scale $k_{\rm tr}$, assumed to lie close to the Planck scale, is presently unknown, and signals the onset of the fixed-point behavior.

We study scenarios with $n\geq 2$ flat extra dimensions \cite{ArkaniHamed:1998rs} with compactification radius $r$, where the fundamental Planck scale $M_{\ast}^{n+2}\!= \frac{M_{\rm Planck}^2}{(2 \pi r)^n}$. A Kaluza-Klein tower of graviton states contributes to the cross section, and has to be summed at the amplitude level.
As for settings with $M_{\ast} \geq 10 \,\rm TeV$, the spacing between Kaluza-Klein modes is small, the sum can be well approximated by an integral. This is UV divergent for $n \geq 2$ when $G_N = \rm const$, see \cite{Cheung:1999ja}. This differs in asymptotic safety, where we employ the scale-identification $k^2 = m^2$, i.e., the fixed-point regime is probed by the heavy Kaluza-Klein modes, whereas the low-lying modes probe the classical gravity regime. Then, using $Z(m)= \frac{G_{\rm Newton}}{G_N(m)}$ yields a finite integral:
\begin{eqnarray}
&{}&\!\int_0^{\infty} \!\!\!dm \frac{m^{n-1}} {Z(m)\! \left(s-m^2\right)}=\! \int_0^{k_{\rm tr}}\!\!\!\!dm \frac{m^{n-1}}{s-m^2}+ k_{\rm tr}^{n+2}\int_{k_{\rm tr}}^{\infty} \!\!\!\!dm \frac{m^{n-1}}{m^{n+2}\!\left(s-m^2\right)}.\label{KK_ints}
\end{eqnarray}.

\begin{figure}[!here]
\includegraphics[width=0.56\linewidth]{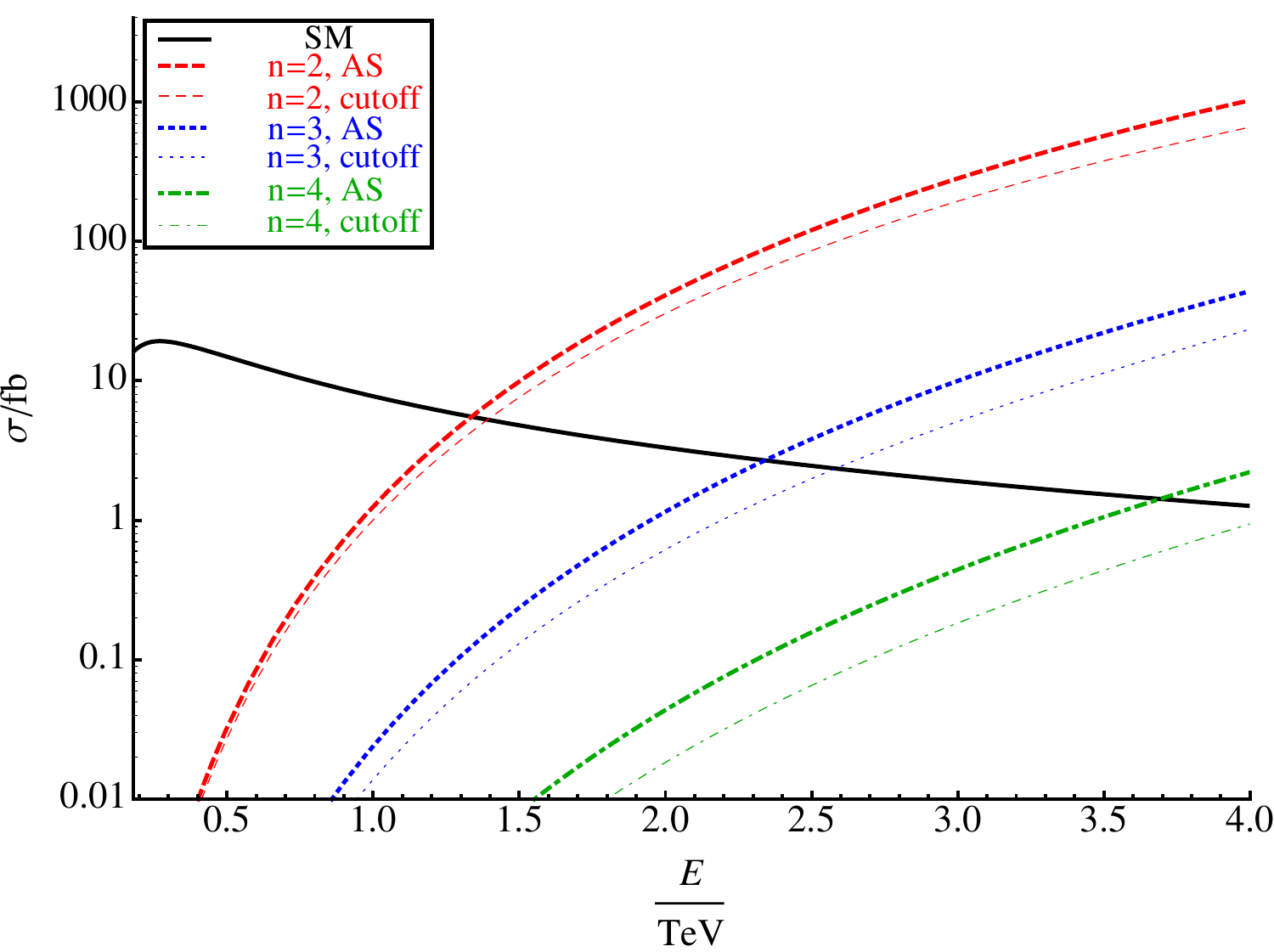}
\caption{\label{gg_SM_AS_n} We plot the cross section ($\pi/6\leq\theta\leq5 \pi/6 $) for 
the standard model (black continuous line), for the cutoff theory (thin lines) with $M_{\ast}= 10\rm TeV$ and for asymptotic safety (thick lines) with $k_{\rm tr}= 10 \rm TeV$. We 
show $n=2$ (red dashed), $n=3$ (blue dotted) and $n=4$ (green dot-dashed). }
 \end{figure}

Fig.~\ref{gg_SM_AS_n} shows our main results \cite{Dobrich:2012nv}: At photon energies of $\sim 1 \, \rm TeV$, the standard-model cross section is measurable at $\sigma \sim 10 \,\rm fb$ and the graviton-induced cross section in our approximation is of the same size, thus being easily detectable. This holds for a fundamental Planck scale $M_{\ast} \sim 10 \, \rm TeV$ and for 2, 3 and possibly even 4 extra dimensions. The result for asymptotic safety clearly differs from the result in the cutoff-theory, due to the extra contribution arising from the fixed-point regime. This allows to experimentally distinguish asymptotic safety from other UV completions.

\section{Conclusions}
Employing sources of high-energy photons, such as an electron collider or laser plasma wake-field acceleration, where photons reach high energies by Compton backscattering off high-energy electrons, allows to test the dimensionality of spacetime and detect possible quantum gravity effects in a very clean setting. Photon energies of 1 TeV suffice to access a fundamental Planck scale $M_{\ast} \sim 10 \, \rm TeV$. Furthermore, the imprints of asymptotically safe quantum gravity can then be clearly distinguished from other quantum gravity models. Thus, near-future experiments can start to shed light on the small-scale structure of spacetime.

\emph{Acknowledgements}
Research at Perimeter Institute is supported by the
Government of Canada through Industry Canada and by
the Province of Ontario through the Ministry of Research
and Innovation.

\bibliographystyle{ws-procs975x65}
%\bibliography{ws-pro-sample}

\end{document}